\documentclass{article}

\usepackage[english]{babel}
\usepackage{amsmath}
\usepackage{mathabx}
\usepackage{booktabs}
\usepackage{natbib}

\usepackage[letterpaper,top=2cm,bottom=2cm,left=3cm,right=3cm,marginparwidth=1.75cm]{geometry}

\usepackage{float}
\usepackage{subcaption}
\usepackage{graphicx}
\usepackage[colorlinks=true, allcolors=blue]{hyperref}

\title{Measuring Opportunity Cost with Stock Lifetime Value}

\author{
  \begin{tabular}[t]{c c c c}
    Geoffrey Decrouez & Tobias Huelden & Paresh Nakhe & Dominik Prugger\\
    Zalando SE & Zalando SE  & Zalando SE & Zalando SE \end{tabular}
}

\begin{document}
\maketitle

\begin{abstract}
Measuring the long-term opportunity cost of interventions remains a critical challenge in e-commerce A/B testing. While strategic levers—such as dynamic pricing, ranking algorithms, and promotional campaigns—trigger shifts in consumer behaviour that persist over months, operational constraints necessitate fast decision-making cycles that are typically limited to weekly experimental windows. Standard metrics like revenue and conversion are inherently short-sighted, biasing decisions toward immediate gains. We introduce Stock Lifetime Value (SLV), a stock-centric metric that captures long-term opportunity cost within short experiments by aggregating expected profit from current inventory through the end of its selling lifecycle. We develop the methodology in the context of fashion e-commerce at Zalando, where stock constraints and seasonal lifecycles make the trade off between short-term and long-term outcomes particularly relevant. SLV aggregates the expected profit from current inventory through the end of its selling lifecycle, providing a way to evaluate interventions against their true profit impact. We discuss three applications: (a) SLV efficiency as a metric for article-level and customer-level A/B tests, validated against realized 18-month lifecycle outcomes; (b) SLV as an optimization target for pricing algorithms, aligning the metric used for measurement with the objective used for decision-making; and (c) a framework for annualizing treatment effects into financial reporting metrics required by business stakeholders. While our empirical setting is fashion retail, the framework applies broadly to any inventory-constrained environment where value decays over time or interventions shift demand across periods.
\end{abstract}

\vspace{0.5em}

\section{Introduction}

A/B tests have become a widely used basis for product development and commercial evidence based decision-making in e-commerce. Decision-makers and experimenters frequently aim to quantify long-term effects that unfold over longer horizons such as months, quarters or years. However, practical business constraints such as traffic costs, temporal opportunity costs and competitive pressure as well as the accelerated pace of decision-making often necessitate short-term experimental windows measured in weeks.  One popular approach to address this issue is to build a model that predicts the long-term impact of a short treatment, using a method based on surrogates \cite{athey2019surrogate}. In this paper, we adopt this approach by estimating the long-term opportunity cost of short-term commercial interventions by introducing a quantity called Stock Lifetime Value (SLV). Following our internal business accounting model, it  represents the total aggregated future profit generated under business-as-usual decisions.

We make three core contributions. First, we introduce the concept of SLV efficiency, a short-term computable metric designed to estimate the opportunity cost of selling perishable inventory. We explain how SLV efficiency complements traditional short-term metrics such as revenue and profit for decision-making and validate its effectiveness on a historical A/B test at Zalando. We provide empirical evidence supporting that the SLV efficiency metric yields robust and consistent decisions when compared against the materialized long-term outcome. Second, we expand the application of SLV beyond the article level measurement problem, detailing how SLV can be used for customer level A/B tests or used as a long-term opportunity cost input for automated pricing algorithms, allowing decision-makers to trade-off short term gains versus long-term costs.
Third, we provide a structured framework for annualizing SLV-based treatment effects from short-term A/B tests. The methodology translates experimental results into the financial reporting framework required by business and finance stakeholders especially relevant to publicly traded companies. 


We develop SLV in the context of a stock-constrained pricing at Zalando, Europe's largest online fashion and lifestyle platforms serving more than 60 million customers across 29 European markets. Zalando operates both as a direct retailer, purchasing and selling inventory under its own account, and as a marketplace connecting third-party brands with consumers.
In 2024, Zalando served approximately 52 million active customers, processing over 250 million orders. It generated €15.3 billion in gross merchandise volume and €10.6 billion in revenue, with an adjusted EBIT of €511 million. As a large wholesaler with multi-billion annual investments in stock, Zalando faces a large challenge to manage under - and overstock risks, with markdown pricing being a key mitigation strategy as well as demand driver. Managing pricing for more than 600,000 products requires algorithmic automation. The fashion retail context—characterized by seasonal lifecycles, uncertain demand, and finite inventory—creates a complex two-sided opportunity cost problem. Conservative discounting risks build-up of overstock and financial write-offs, while aggressive discounting sacrifices margin on units that could have sold at higher prices. This economic tension is compounded by organizational frictions: trading teams often maximize for tangible and measurable short-term revenue, whereas leadership requires credible estimates of long-term profitability for planning and growth. Consequently, this environment of high volumes and perishable inventory provides an ideal setting to analyze long-term measurement challenges in e-commerce experimentation.

The framework presented in this paper has become integral to how experimental pricing initiatives are evaluated and communicated at Zalando. Since its introduction, the approach has been applied to numerous pricing A/B tests, with individual initiatives yielding projected annual impacts in the double-digit million euro range for both revenue and profit. These estimates feed directly into quarterly business reviews and earnings reports. Moreover, the quantitative insights have strengthened the business case for continued investment in algorithmic pricing and the experimentation infrastructure.
While our empirical setting is fashion retail, the SLV framework applies more broadly to any business environment where inventory is constrained, value decays over time, or interventions shift demand across periods rather than simply increasing it. This includes other retail verticals with seasonal or perishable goods, travel and hospitality with fixed capacity, event ticketing, and any marketplace where supply is limited and timing affects value. The core requirement is that selling today has an opportunity cost—foregoing the option to sell the same unit later under potentially different conditions.

Long-term measurement is cited in \cite{gupta2019top} as one of the most challenging problems in online experimentation: short-term A/B test metrics often fail to predict long-term performance. The challenge to predict long-term performance from short term metrics is illustrated in \cite{hohnhold2015focusing} in the context of advertising at Google, or in \cite{kohavi2012trustworthy} in a Bing search experiment, where short-term metrics like "queries per user" or "revenue per user" should not be used as the sole Overall Evaluation Criterion (OEC) for search/ad experiments. We adopt a similar position in this paper, where we advice to report on SLV efficiency in addition to short-term revenue and profit outcomes. 
Our approach can be expressed within the surrogate framework, introduced by \citep{prentice1989surrogate}. The surrogacy assumption requires the surrogate to fully mediate the path between short-term treatment and long-term outcome. \cite{athey2019surrogate} builds on this framework by combining multiple short-term outcomes into a single "surrogate index". Technology companies have extensively adopted and refined surrogate methods for long-term causal inference, for example, at Netflix \cite{zhang2024evaluating}, \cite{bibaut2023longterm}, at LinkedIn \cite{duan2021linkedin}, or at Spotify in the presence of latent confounders \cite{vangoffrier2023estimating}. However, applications to date have focused mainly on long-term customer-centric outcomes such as customer lifetime value (CLV) \cite{fader2005counting}. Less work has been done to quantify the impact of commercial levers on standard performance indicators such as yearly revenue and profitability. We fill this gap by introducing a stock-centric metric based on the surrogate's framework.



 
The remainder of this paper is organized as follows. Section \ref{section:methodology} develops the methodology: we formally define SLV and its normalized version nSLV, and describe our approach to forecasting SLV. Section \ref{section:applications} presents various applications of SLV. In \ref{subsection:measurement}, we first demonstrate how SLV improves measurement efficiency in A/B tests by enabling reliable long-term inference from short experimental windows. Section \ref{subsection:customerslv} connects the article based SLV metric with customer based A/B tests. We then illustrate in \ref{subsection:optimal_discounting}, how SLV serves as an optimization target for pricing algorithms, framing markdown decisions as an explicit trade-off between short-term revenue and long-term profit. Finally, we show in \ref{subsection:annualization} how experimental results can be annualized for financial reporting and business impact communication. Section \ref{section:conclusion} concludes with a discussion of limitations and directions for future work.

\section{Methodology}
\label{section:methodology}

\subsection{Stock Lifetime Value Definition}
\label{sub:slvdef}

For retailers managing perishable or seasonal goods, the core operational challenge is the efficient liquidation of their stock. At Zalando, the majority of inventory is procured prior to the start of a 6-month commercial cycle (Spring/Summer and Autumn/Winter). Because within-season replenishment is limited, pricing and ranking interventions operate on a fixed amount of stock. Consequently, every unit sold today represents a trade-off: immediate liquidity versus the potential for higher-margin realization later in the season.

To quantify this trade-off, we define SLV as the expected aggregate profit generated by the current units in stock until the end of their selling lifecycle. We adopt a sunk-cost perspective typical in dynamic pricing, assuming that wholesale purchase prices are irreversible. Under this assumption, the objective is to maximize Marginal Profit (MP)—the total revenue minus variable operational expenses, such as fulfillment, payment processing, and logistics costs.

The precise definition of SLV is modular, allowing it to adapt to the accounting scope of the business. In the context of Zalando, the aggregation of MP accounts for multi-channel liquidation strategies. This includes secondary channels such as Zalando Lounge, which mainly serves as a terminal clearance path for unsold inventory, with high discounting. By incorporating these downstream channels into SLV, we ensure that experimental interventions are evaluated not just by their immediate on-platform performance, but by their impact on the total recovery value of the stock.

Let $N_{i,t}$ denote the total number of units $i$ sold on day $t$ (after returns), $\text{NMV}_{i,t}$ the Net Merchandising Value realised from the sales, $\text{MP}_{i,t} = \text{NMV}_{i,t} - \text{IncrementalCosts}_{i,t}$ the associated total marginal profit, and $S_{i,t}$ the number of units $i$ in stock at the start of day $t$. For a given date $t_0$, we define $\text{SLV}_{i, t_0}(S_{i,t_0})$ as the total marginal profit made after $t_0$, conditionally on the stock levels $S_{i,t_0}$,
\begin{equation}\label{slvdef}
\text{SLV}_{i, t_0}(S_{i,t_0}) = \sum_{t\geq t_0} \text{MP}_{i,t}\mathbf{1}\left(\sum_{s=t_0}^t N_{i,s} \leq S_{i,t_0}\right)\,,
\end{equation}
where $\mathbf{1}(\cdot)$ is the indicator function, equal to 1 when the cumulative sales $\sum_{s=t_0}^t N_{i,s}$ up to time $t$ is less than the stock levels at time $t_0$, and 0 otherwise. This is illustrated in Figure \ref{fig:zalando_SLV}, and an example calculation is provided in Appendix \ref{appendix:slv}. SLV disregards future replenishments of stock that are potentially based on past performance: goods that perform well can be restocked, leading to additional incremental marginal profit not captured by $\text{SLV}_{i, t_0}(S_{i,t_0})$. The actual materialized SLV is calculated using historical sales data covering the entire lifecycle, typically 18 months, of our assortment.

\begin{figure}[t]
    \centering
    \includegraphics[width=12cm]{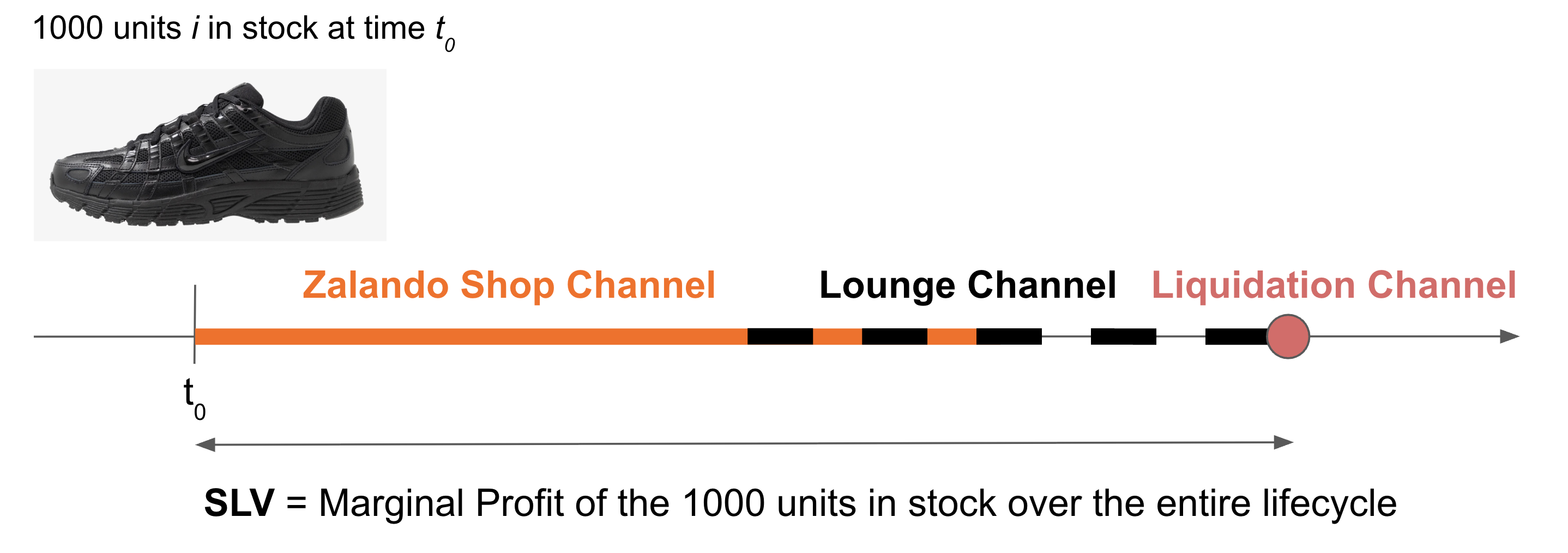}
    \caption{$\text{SLV}_{i, t_0}(S_{i,t_0})$ represents the total marginal profit generated from the $S_{i,t_0}=1000$ units of sneakers $i$ in stock at time $t_0$, via all the Zalando channels : shop, lounge, and liquidation.}
    \label{fig:zalando_SLV}
\end{figure}

We introduce another quantity called normalized SLV (nSLV), defined as $\text{SLV}_{i, t_0}(S_{i,t_0})$ normalized by the stock value at $t_0$. Denoting $\text{pp}_{i}$ the purchase price of unit $i$ at the start of the season, the stock value at $t_0$ is $\textrm{SV}_{i,t_0} = S_{i,t_0}\cdot \text{pp}_{i}$, so that 
$$
\text{nSLV}_{i, t_0} = \frac{\text{SLV}_{i, t_0}(S_{i,t_0})}{\textrm{SV}_{i,t_0}}
$$
represents the average marginal profit per euro invested at $t_0$. Normalizing SLV significantly improves interpretability and facilitates comparisons across different groupings of articles. For instance, $\text{nSLV}_{i, t_0} = 1$ indicates a break-even scenario: over the article's remaining lifetime, the net earnings from available stock exactly equal the initial purchase investment, resulting in zero net profit. Consequently, articles with $\text{nSLV}_{i, t_0} > 1$ are identified as profitable, with $\text{nSLV}_{i, t_0}-1$ representing the average profit margin. 

\begin{figure}[t]
    \centering
    \includegraphics[width=14cm]{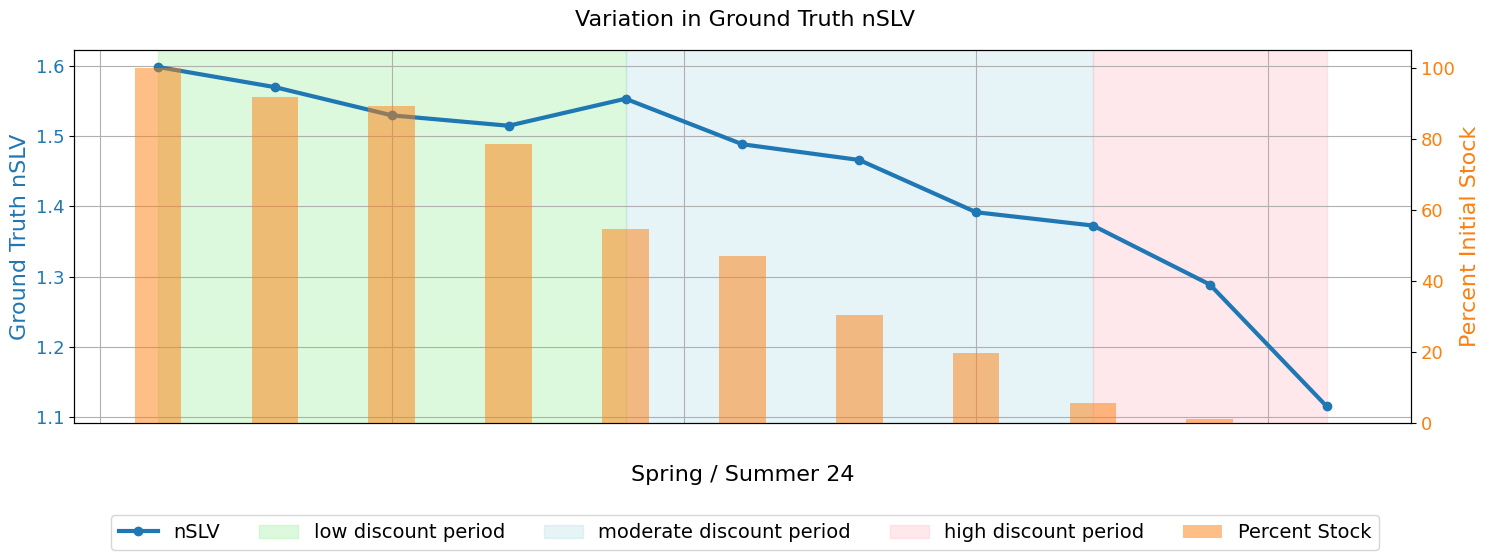}
    \caption{Variation in nSLV (blue) for a specific model of training shoes over the course of a season and the corresponding changes in stock (orange). Low to high discounting periods are also provided.}
    \label{fig:RV_Variation}
\end{figure}

Figure \ref{fig:RV_Variation} illustrates the typical variation of nSLV in the context of Zalando for a particular model of training shoes. During the initial weeks of the lifecycle, nSLV remains relatively high, indicating high margins, corresponding to high seasonal demand and low discounting. Towards the end of the season, aggressive discounting policies are required to liquidate the stock, typically leading to nSLV values reaching the vicinity of 1.

\vspace{1em} 
\noindent \textbf{Remark.} The SLV concept is distinct from another established industry metric, Customer Lifetime Value (CLV), which has been around for decades to measure the total profit a customer is expected to generate over their entire relationship with the company. Although SLV is a stock-centric entity, in Section \ref{subsection:customerslv} we explain how SLV-based metrics can be defined at the customer level, allowing optimization and measurement of targeted customer actions, such as personalized vouchers, by linking customer behavior directly to the inventory outcomes.

\subsection{nSLV Forecast}
\label{section:rvforecast}

\begin{figure}[t]
    \centering
    \includegraphics[width=15cm]{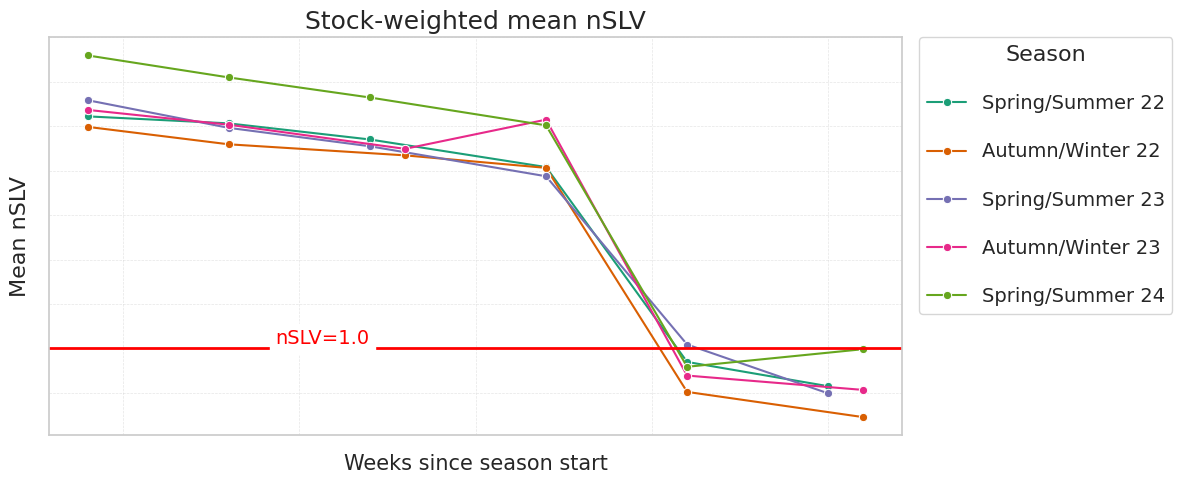}
    \caption{Stationarity in operational lifecycle and pricing processes within Zalando captured as variation in nSLV (y-axis) over the course of 5 consecutive seasons. The trajectory of nSLV follows a predictable pattern each season: starting with a high nSLV and decreasing gradually and falling just below 1, the break-even point, towards the end of the season. The y-axis is anonymised to keep the order of magnitude confidential at an aggregated level.}
    \label{fig:mean_nSLV}
\end{figure}

The applications we discuss in Section \ref{section:applications} need an SLV forecaster with the ability to learn how marginal profit is impacted by demand patterns, sales velocity, stock levels, among other factors that can vary over the year. For robust predictions, a high-degree of structural stationarity in the underlying data generating processes governing SLV is needed. In the context of fashion industry, the regular and repeating operational lifecycles offer an ideal environment. The typical fashion lifecycle consists of three phases: in the premium phase at the beginning of a season, margins are maximized by leveraging fresh assortments targeted at fashion-aware consumer segments; mid-season dynamics characterized by incremental price adjustments used to align sales velocity with inventory targets followed by a stock-clearance phase characterized by aggressive discounts at season end. This is illustrated in Figure \ref{fig:mean_nSLV}, where we observe regular $nSLV$ patterns across seasons. In addition, at Zalando, individual drivers of price variation—including algorithmic discounts, localized voucher campaigns and competitive price-matching—jointly characterize the stationarity in the pricing processes. While localized, time-limited variations may occur in response to exogenous factors, we treat these as noise from a forecaster-design perspective. We refer to this stationarity in the operational lifecycle and the pricing processes together as the ``business-as-usual" assumption.


We use $\text{nSLV}_{i, t_0}$ as the target for the forecaster. This choice is based on the observation that the SLV metric exhibits high variance and a long-tailed distribution. Such characteristics hamper the model's stability and convergence, potentially resulting in a non-uniform focus across the dataset that disproportionately weights articles with high absolute errors. Using nSLV—a low-variance metric—as the target, we decouple the features from the target, forcing the model to capture key patterns that drive profit margins uniformly across all articles.



The training features, denoted $\mathbf{X}_{i, t_1}$ for a historical date $t_1$, are constructed from approximately 15 million article-level data points spanning two years of historical data. These features fall into four primary categories:

\begin{itemize}
    \item \textbf{Static Categorical Attributes}: Broad article characteristics, including material (e.g., Textile), category (e.g., Men’s straight-fit Jeans), brand, and price cluster.
    \item \textbf{Article-Specific Seasonality}: Features capturing the season type and the article's current position within its demand cycle, accounting for the regular variations resulting from voucher campaigns.
    \item \textbf{Commercial performance}: Multi-granular sales metrics tracked throughout the article's lifecycle, such as profit margin, sales velocity, and stock cover.
    \item \textbf{Stock dynamics}: Inventory levels and delivery schedules, including the timing of potential reorders.
\end{itemize}

The model utilizes multi-output regression via the vector-leaf variant of XGBoost to generate simultaneous predictions for the start and end of a specific interval. This multi-output architecture is particularly effective at preserving the correlations between these two points—a property that is crucial for the A/B testing framework discussed in the following section. Furthermore, the training procedure is designed to replicate the temporal context available during inference by utilizing separate training runs for each distinct interval. Once $\widehat{\text{nSLV}}_{i, t_0}$ is available, the SLV prediction is computed using
$$
\widehat{\text{SLV}}_{i, t_0}(\text{S}_{i,t_0})= \widehat{\text{nSLV}}_{i, t_0} \cdot \textrm{SV}_{i,t_0}\,.
$$

\begin{figure}[t]
    \centering
    \includegraphics[width=15cm]{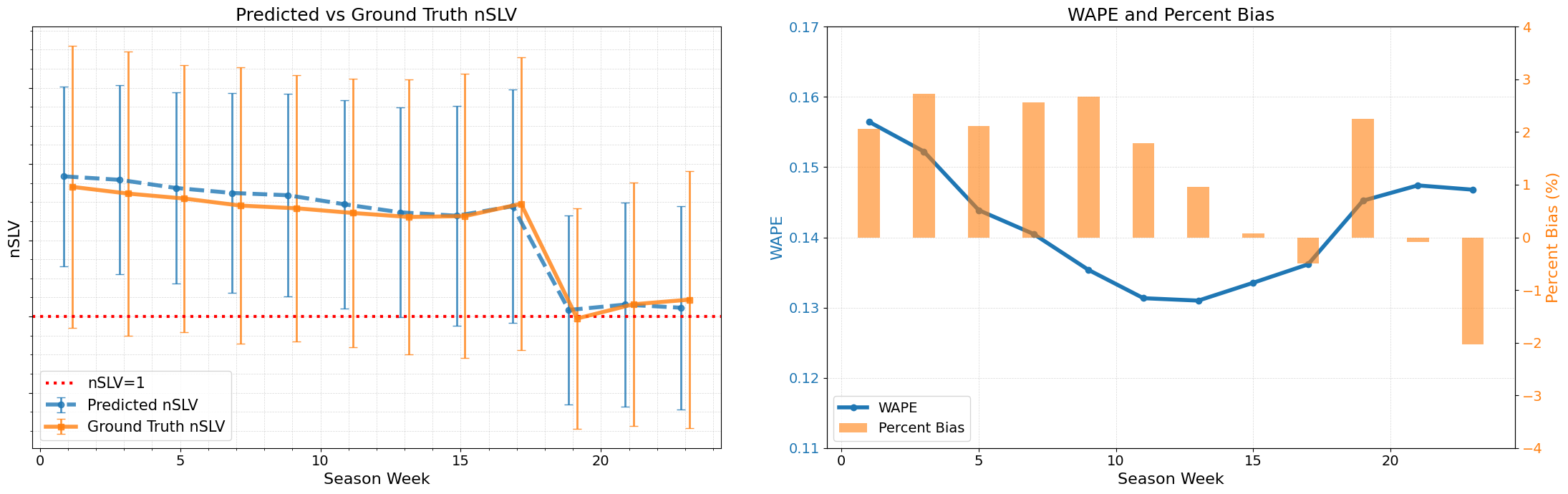}
    \caption{(left) Average $\text{nSLV}_{t_0}$ and $\widehat{\text{nSLV}}_{t_0}$ at Zalando over the full seasonal assortment. The dotted red line represents the break even line $\text{nSLV}_{i, t_0}=1$. The x-axis represents the number of weeks since season start. The y-axis is anonymised to keep the order of magnitude confidential at an aggregated level. (Right) Accuracy metrics WAPE and bias computed in a backtesting environment.}
    \label{fig:zalando_RV}
\end{figure}

Figure \ref{fig:zalando_RV} presents the forecast accuracy, evaluated in a backtesting environment. The assortment comprises of all the seasonal articles available for sale on the website and with an evaluation horizon spanning the entire season. The model is evaluated in terms of the prediction accuracy of SLV at a given point in time. Specifically, we use the Weighted Absolute Percentage Error (WAPE)
$$\text{WAPE}  = \frac{\sum\limits_{i} \lvert  \widehat{SLV}_{i, t_0} - SLV_{i, t_0} \rvert}{\sum\limits_{i} SLV_{i, t_0}}$$
and 
$$\text{Bias}  = \frac{\sum\limits_{i} \left( \widehat{SLV}_{i, t_0} - SLV_{i, t_0} \right) }{\sum\limits_{i} SLV_{i, t_0}} \times 100, $$
measuring the percentage error in the total estimated SLV. To evaluate the model at date $t_0$, we use the features and ground-truth nSLV from equivalent points in time across the previous four seasons. This evaluation is repeated at two-week intervals throughout the entire season. The left plot of Figure \ref{fig:zalando_RV} depicts the average seasonal trend for the stock-weighted ground-truth nSLV alongside its corresponding forecast. Error bars represent the standard deviation of the distributions. The dip in nSLV observed in week 19 reflects the onset of end-of-season sales and increased discount depths. While the right plot indicates that predicted SLV has an average article-level error of 15\%, the left plot demonstrates that predictive accuracy improves significantly when the data is aggregated.

\section{Applications}
\label{section:applications}


\subsection{Article-based A/B Tests}
\label{subsection:measurement}


\subsubsection{SLV efficiency definition}
\label{subsection:slvefficiency}

At Zalando, we regularly measure the efficiency of new pricing policies using A/B tests, also known as randomized controlled trials. Randomization occurs on the eligble assortment at an article-cluster level rather than on a customer level as described in detail in \cite{Huelden2024}. We cluster similar articles to mitigate substitution effects that would otherwise violate the Stable Unit Treatment Value Assumption (SUTVA). However, reporting test results of short-term metrics such as revenue and profit gives us an incomplete picture as they do not take into account long term opportunity costs: a pricing policy that increases discounts may yield a positive estimate on immediate revenue, but the stock sold early in the season might have been sold later at a higher margin, rendering the policy ultimately inefficient. To address this, we construct a new metric, SLV efficiency, which incorporates long term stock dynamics and directly weighs the short term impact against its long term opportunity cost.

For a unit $i$, and a test conducted between $t_1$ and $t_2$, let $Z_i\in\{0,1\}$ denote its binary treatment allocation. Let $\mathcal{T}$ denote the set of units in the treatment group ($Z_i=1$), and $\mathcal{C}$ denote the set of units in the control group ($Z_i=0$). The number of units in each group is given by $n_\mathcal{T}=\sum_i Z_i$ and $n_\mathcal{C}=\sum_i (1-Z_i)$. Recall that $\text{SLV}_{i, t_1}(S_{i, t_1})$ denotes the SLV of a unit $i$ at $t_1$, conditionally on the number of units in stock $S_{i, t_1}$. We introduce $\text{SLV}_{i, t_2}(S_{i, t_2}^{\star})$ with stock levels at $t_2$ taken as $S_{i, t_2}^{\star} = max(0, S_{i, t_1} - \sum_{t=t_1}^{t_2} N_{i,t})$ instead of $S_{i, t_2}$, representing the adjusted stock levels given that we sold $min(\sum_{t=t_1}^{t_2} N_{i,t}, S_{i, t_1})$ between $t_1$ and $t_2$. Note that $S_{i, t_2}$ and $S_{i, t_2}^{\star}$ differ in the event of non anticipated returns by customer before the start of the test, or non anticipated restocking during the test, leading to more sales between $t_1$ and $t_2$ than the observed stock levels at $t_1$. 

We now introduce a quantity called SLV consumed, representing the business-as-usual marginal profit value consumed between $t_1$ and $t_2$, for the $min(\sum_{t=t_1}^{t_2} N_{i,t}, S_{i, t_1})$ units sold during the experiment. Moving forward, it is convenient for identification to use the potential outcome notation. Let:
\begin{itemize}
\item $\text{SLV}_{i, t_1}(S_{i, t_1})(0, 0)$ defined as in (\ref{slvdef}), be the potential outcome equal to the total marginal profit generated from the $S_{i, t_1}$ units in stock at $t_1$, for a never treated unit $i$, under ``business-as-usual" pricing.
\item $\text{nSLV}_{i, t_1}(0,0) = \text{SLV}_{i, t_1}(S_{i,t_1})(0,0) / (S_{i,t_1}\cdot \text{pp}_{i})$
\item $\text{SLV}_{i, t_2}(S_{i, t_2}^{\star})(j, 0)$ defined as in (\ref{slvdef}), be the potential outcome equal to the total marginal profit generated from the $S_{i, t_2}^\star$ units in stock at $t_2$, for a unit $i$ that received treatment $j\in\{0,1\}$ between $t_1$ and $t_2$, and no treatment after $t_2$ (``business-as-usual").
\item $\text{nSLV}_{i, t_2}(j,0) = \text{SLV}_{i, t_2}(S_{i,t_2}^\star)(j,0) / (S_{i,t_2}^\star\cdot \text{pp}_{i})$
\end{itemize}
The SLV consumed value, which represents the marginal profit consumed by the experiment's sales against the business-as-usual opportunity cost, is then defined as the difference between these two potential outcomes
$$
\text{SLVcons}_{i, t_1, t_2}(j) = \text{SLV}_{i, t_1}(S_{i, t_1})(0,0)-\text{SLV}_{i, t_2}(S_{i, t_2}^{\star})(j,0)\,,
$$ 
where we drop the explicit dependence on the stock levels $(S_{i, t_1}, S_{i, t_2}^{\star})$ in the notation.
For a unit $i$ receiving treatment $j=1$, the potential outcome $\text{SLV}_{i, t_1}(S_{i, t_1})(0,0)$ is never observed, and neither is $\text{SLVcons}_{i, t_1, t_2}(1)$. In practice, we estimate these two terms with the forecast introduced in Section \ref{section:rvforecast}. SLV consumed is then compared with the short-term marginal profit generated during the experiment, leading to the SLV efficiency metric 
\begin{equation}\label{eq:slveff}
\text{SLVeff}_i(S_{i, t_1})(j) = \sum_{t=t_1}^{t_2} \text{MP}_{i,t} \mathbf{1}\left(\sum_{s=t_1}^t N_{i,s} \leq S_{i,t_1}\right) - \text{SLVcons}_{i, t_1, t_2}(j)\,.
\end{equation}
At the end of the test, we report the difference $\widebar{\Delta \text{SLVeff}}=\widebar{\text{SLVeff}}_{trt} - \widebar{\text{SLVeff}}_{ctl}$, where 
\begin{align}
\widebar{\text{SLVeff}}_{trt} &=  \frac{1}{n_\mathcal{T}}\sum_{i } Z_i\cdot \text{SLVeff}_i(S_{i, t_1})(1)  \notag \\
\widebar{\text{SLVeff}}_{ctl} &= \frac{1}{n_\mathcal{C}}\sum_{i} (1-Z_i)\cdot \text{SLVeff}_i(S_{i, t_1})(0)\,. \notag
\end{align}
A positive $\widebar{\Delta \text{SLVeff}}$ indicates that a new pricing policy yields a more efficient usage of stock compared to a “business-as-usual” scenario, while negative SLV efficiency results indicate less efficient policies.

\begin{figure}[t]
    \centering
    \includegraphics[width=13cm]{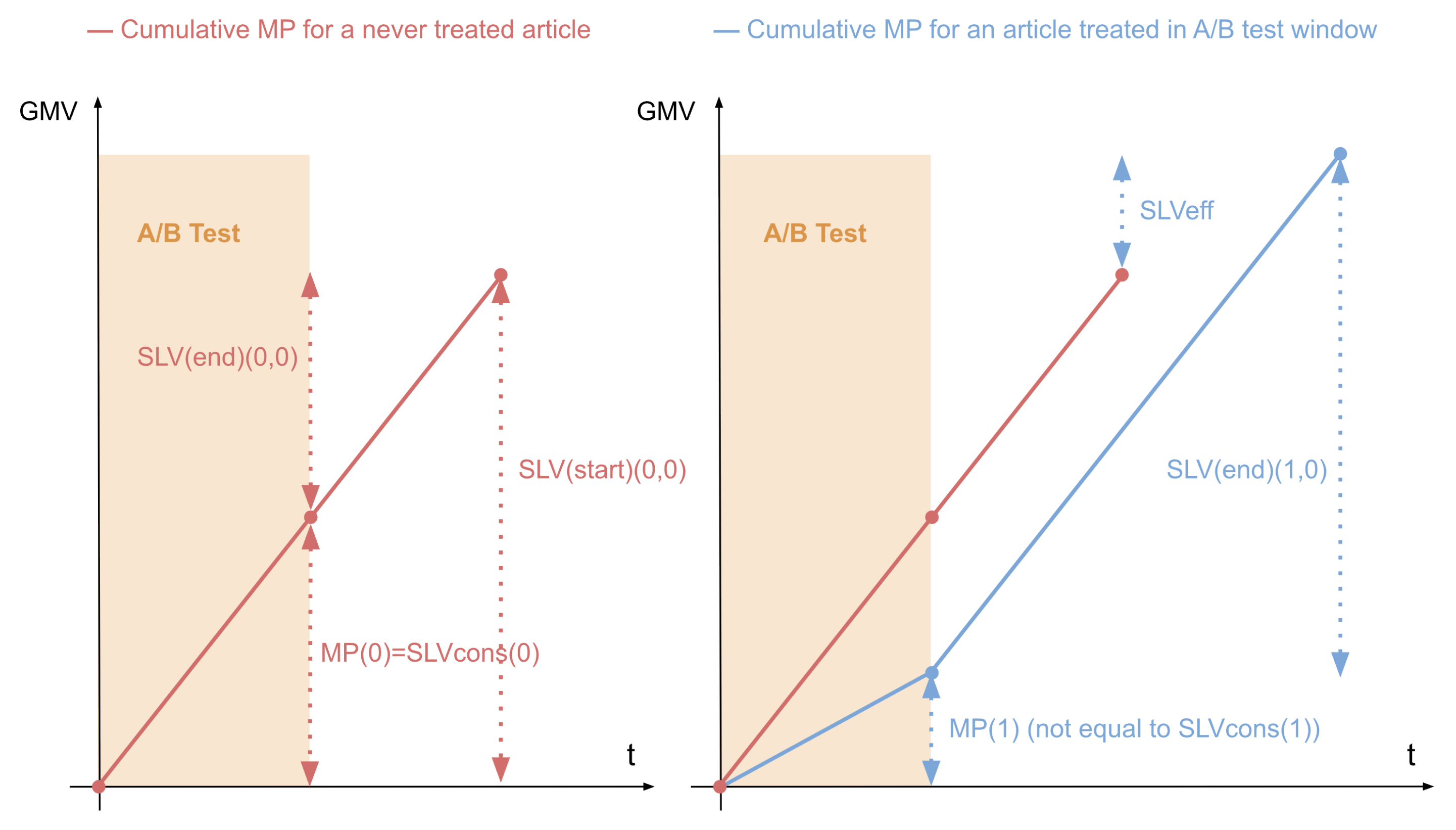}
    \caption{Illustration of SLV efficiency. Both plots show the cumulative marginal profit of an article over the course of its lifetime, conditionally on the number of items in stock at the start of an A/B test. The left plot show counterfactual values for an item never treated. In this case, the marginal profit cumulated during the test is equal to SLV consumed, and the unit-level SLV efficiency is 0. On the right hand side, we show the trajectory of the same item if it received the treatment in the short A/B test window. Assuming the demand is lower during the test, stock levels are higher if the unit was treated. The additional stock is sold for a longer time period. Assuming the item does not need to be heavily discounted to be liquidated, the additional stock at test end brings overall more marginal profit than if not treated in the A/B test window, leading to a positive SLV efficiency.}
    \label{fig:zalando_SLVeff}
\end{figure}

As noted before, $\text{SLVcons}_{i, t_1, t_2}(1)$ is not observed for a treated unit and is estimated using the forecast of Section \ref{section:rvforecast}. 
With forecasts $\widehat{\text{nSLV}}_{i, t_1}$ and $\widehat{\text{nSLV}}_{i, t_2}$ of $ \text{nSLV}_{i, t_1}(0,0)$ and $\text{nSLV}_{i, t_2}(0,0)$, put
\begin{align}
\widehat{\text{SLV}}_{i, t_1}(S_{i, t_1}) &=  \widehat{\text{nSLV}}_{i, t_1} \cdot \text{pp}_{i} \cdot \text{S}_{i,t_1}  \notag \\
\widehat{\text{SLV}}_{i, t_2}(S_{i, t_2}^{\star}) &= \widehat{\text{nSLV}}_{i, t_2} \cdot \text{pp}_{i} \cdot S_{i, t_2}^{\star} \notag \\
\widehat{\text{SLVcons}}_{i, t_1, t_2} &= \widehat{\text{SLV}}_{i, t_1}(S_{i, t_1}) - \widehat{\text{SLV}}_{i, t_2}(S_{i, t_2}^{\star}) \notag
\end{align}
and
\begin{alignat}{2}
\widehat{\text{SLV}}_{trt, t_1} &=  \frac{1}{n_\mathcal{T}}\sum_{i } Z_i\cdot \widehat{\text{SLV}}_{i, t_1}(S_{i, t_1}) & \quad \widehat{\text{SLV}}_{trt, t_2} &= \frac{1}{n_\mathcal{T}}\sum_{i } Z_i\cdot \widehat{\text{SLV}}_{i, t_2}(S_{i, t_2}^\star) \notag \\
\widehat{\text{SLV}}_{ctl, t_1} &= \frac{1}{n_\mathcal{C}}\sum_{i} (1-Z_i)\cdot \widehat{\text{SLV}}_{i, t_1}(S_{i, t_1})\,. & \quad \widehat{\text{SLV}}_{ctl, t_2} &= \frac{1}{n_\mathcal{C}}\sum_{i } (1-Z_i)\cdot \widehat{\text{SLV}}_{i, t_2}(S_{i, t_2}^\star)\notag
\end{alignat}
and similarly for SLV consumed and efficiency. We work under the assumption that $\text{nSLV}_{i, t_2}(1,0)=\text{nSLV}_{i, t_2}(0,0)$, stating profit after $t_2$ is independent of sales between $t_1$ and $t_2$, given the stock levels at the end of the test. This assumption is revisited in Section \ref{subsection:surrogates}, where we link our approach with surrogates.


\vspace{1em} 
\noindent \textbf{Remark.} Randomisation ensures that the average SLV at the start of the experiment is approximately balanced, $\widehat{\text{SLV}}_{trt, t_1} \approx \widehat{\text{SLV}}_{ctl, t_1}$, so that
\begin{align}
\widehat{\Delta \text{SLVeff}} &= (\widebar{\text{MP}}_{trt}-(\widehat{\text{SLV}}_{trt, t_1}-\widehat{\text{SLV}}_{trt, t_2}))- (\widebar{\text{MP}}_{ctl}-(\widehat{\text{SLV}}_{ctl, t_1}-\widehat{\text{SLV}}_{ctl, t_2}))\notag \\
&\approx (\widebar{\text{MP}}_{trt}+\widehat{\text{SLV}}_{trt, t_2})- (\widebar{\text{MP}}_{ctl}+\widehat{\text{SLV}}_{ctl, t_2})\,, \label{deltaslveff}
\end{align}
showing that $\widehat{\Delta \text{SLVeff}}$ measures the incremental value of the short-term treatment over the entire lifetime, given that we revert back the treatment to business-as-usual at the end of the experiment. A pricing policy that sells too quickly during the test window leads to a lower value ($\widehat{\text{SLV}}_{trt, t_2} < \widehat{\text{SLV}}_{ctl, t_2}$), which can result in a negative $\widehat{\Delta \text{SLVeff}}$, despite generating more short-tertm marginal profit ($\widebar{\text{MP}}_{trt} > \widebar{\text{MP}}_{ctl}$). This is illustrated in Figure \ref{fig:zalando_SLVeff} for a positive SLV efficiency. 

\vspace{1em}

Finally, we introduce a relative SLV efficiency metric, which is the key quantity that we use in Section \ref{subsection:annualization} for annualisation: 
$$\frac{\widehat{\Delta \text{SLVeff}}}{\widehat{\text{SLVcons}}_{trt, t_1, t_2}}\,.$$
The denominator, $\widehat{\text{SLVcons}}_{trt, t_1, t_2}$ is taken as the reference value since it represents the opportunity costs of the consumed inventory during the A/B test. 
The forecast bias on SLV efficiency in the control group is $\widebar{\text{MP}}_{ctl}-\widehat{\text{SLVcons}}_{ctl, t_1, t_2}$. Assuming a constant bias between the treatment and control group, we consider a bias-corrected version of the denominator using $\widehat{\text{SLVcons}}_{trt, t_1, t_2}+(\widebar{\text{MP}}_{ctl}-\widehat{\text{SLVcons}}_{ctl, t_1, t_2})$, leading to
$$
\widehat{\text{relSLVeff}} = \dfrac{\widebar{\text{MP}}_{trt}-\widebar{\text{MP}}_{ctl}-\widehat{\Delta \text{SLVcons}}_{t_1, t_2}}{\widebar{\text{MP}}_{ctl}+\widehat{\Delta \text{SLVcons}}_{t_1, t_2}}\,,
$$
where $\widehat{\Delta \text{SLVcons}}_{t_1, t_2}= \widehat{\text{SLVcons}}_{trt, t_1, t_2} - \widehat{\text{SLVcons}}_{ctl, t_1, t_2}$. The relative SLV efficiency adjusts the short-term relative lift on marginal profit with the observed difference in SLV consumed. Pricing policies that make efficient use of the stock have $\widehat{\Delta \text{SLVcons}}_{t_1, t_2}<0$, leading to an inflated short-term uplift, reflecting our confidence to sell the remaining additional stock at a later point in the season.

\subsubsection{Link with Surrogates}
\label{subsection:surrogates}
 
Let $\text{SLV}_{i, t_1}(S_{i,t_1})(j,0)$ denote the SLV potential outcome for unit $i$, given that $i$ receives treatment $j$ between $t_1$ and $t_2$ and reverted to ``business-as-usual" after $t_2$. Equation (\ref{deltaslveff}) shows that $\widehat{\Delta \text{SLVeff}}$ estimates the Average Treatment Effect (ATE) of the long-term marginal profit
$$
\tau_{SLV} = \mathbf{E}(\text{SLV}_{i, t_1}(S_{i,t_1})(1,0)-\text{SLV}_{i, t_1}(S_{i,t_1})(0,0)\, | \, S_{i,t_1})\,,
$$
conditionally on the stock levels $S_{i,t_1}$ at the start of the experiment. 
\cite{athey2019surrogate} propose a surrogate framework for estimating the long-term impact of a short-term treatment such as $\tau_{SLV}$. The framework requires to find a set of values, called surrogates, such that the treatment only affects the long-term outcome through the surrogates. In Section \ref{subsection:slvefficiency}, we work under the assumption that the marginal profit generated during the experiment, together with the stock value at the end of the experiment ($(\{\text{MP}_{i,t}\}_{t=t_1}^{t_2}, \text{pp}_{i}\cdot S_{i,t_2}^\star)$) fully mediate the causal path between the short-term treatment $Z_i$ and the long-term outcome $\text{SLV}_{i, t_1}(S_{i, t_1})$, leading to the following causal diagram:
$$
Z_i \xrightarrow{} (\{\text{MP}_{i,t}\}_{t=t_1}^{t_2}, \text{pp}_{i}\cdot S_{i,t_2}^\star) \xrightarrow{} \text{SLV}_{i, t_1}(S_{i, t_1}) \,.
$$
In particular, note that $\text{SLV}_{i, t_1}(S_{i, t_1})= \sum_{t=t_1}^{t_2} \text{MP}_{i,t} \mathbf{1}\left(\sum_{s=t_0}^t N_{i,s} \right) + \text{SLV}_{i, t_2}(S_{i, t_2}^{\star})$ is a sum of two terms: the short-term marginal profit observed during the experiment (the first surrogate), and the remaining profit generated after $t_2$ based on the stock levels $S_{i, t_2}^{\star}$ that materialise at the end of the experiment (the second surrogate). Similarly to \cite{athey2019surrogate}, we use historical data to estimate $\text{SLV}_{i, t_2}(S_{i, t_2}^{\star})$,
$$
\widehat{\text{SLV}}_{i, t_2}(\text{S}_{i,t_2}^\star)= \widehat{\text{nSLV}}_{i, t_1} \cdot \text{pp}_{i} \cdot S_{i,t_2}^\star = \hat{f}(\mathbf{X}_{i, t_1}) \cdot \text{pp}_{i} \cdot S_{i,t_2}^\star = g(\mathbf{X}_{i, t_1}, S_{i,t_2}^\star)\,,
$$
where $\mathbf{X}_{i, t_1}$ are the pre-test features, independent of the treatment. The resulting estimator of the long-term outcome $\text{SLV}_{i, t_1}(S_{i, t_1})$ is a linear combination of the two surrogates, resulting in a simple and easy to interpret surrogate index.

The surrogate framework requires three assumptions to identify $\tau_{SLV}$. The first assumption relies on an unconfounded treatment assignment, which is guaranteed in our setting since the articles are randomized. The second is the fundamental surrogacy assumption. It states that, conditional on the marginal profit realized during the experiment and the stock levels remaining at the experiment's end, the assignment of the treatment discount does not affect the long-term outcome. For this to hold, the new pricing policy must not alter the demand curve for the remainder of the product lifecycle. This implies that customers must not remember the recently displayed discount in a way that delays or alters their future purchases. Although the lowest prices must be displayed for a period of at least 30 days under the European Omnibus directive, a typical article lifecycle at Zalando spans up to 18 months, making this assumption plausible in our context. We evaluate this assumption in a practical use case in the section \ref{subsubsection:illustration} below. The third assumption requires that the experimental and observational samples are comparable. For the SLV framework, this means that the historical sales data used to train the nSLV forecast model is representative of the current business-as-usual operating conditions, as illustrated in Figure \ref{fig:mean_nSLV}.

\subsubsection{Illustration}
\label{subsubsection:illustration}
The estimated SLV efficiency (\ref{deltaslveff}) is based on an SLV forecast at the end of the test, under the assumption that we revert back to business-as-usual. To validate (\ref{deltaslveff}), we examine an experiment where we can compare SLV efficiency estimates against materialised SLV values over the full product lifecycle. This requires an experiment that (a) produced a statistically significant effect on SLV efficiency, (b) was not rolled out to production, and (c) has sufficient time elapsed to observe the complete lifecycle of the participating assortment. These criteria substantially limit the candidate pool consisting of our typical pricing tests: many experiments in our pipeline yield statistically insignificant SLV efficiency estimates, and almost all experiments with statistically significant and positive estimates are rolled out after conclusion.

We, therefore, focus on a special reactivation experiment designed to test whether slow-moving inventory could be stimulated through aggressive pricing interventions. Conducted between July 24 and August 6, 2024, the experiment applied a 20 percentage point discount increase to 9,705 articles representing approximately 8.5 million euro in stock value across the largest 13 European markets, affecting approximately 126,000 article-country pairs. The experimental design employed clustered randomization, with strict price upload enforcement. Each article was either in the same treatment or control group across all the tested markets.
All articles were pre-selected as slow movers based on sales velocity and inventory age criteria.

In a first step, we evaluate the plausibility of the surrogacy assumption $Z_i \xrightarrow{} \text{pp}_{i}\cdot S_{i,t_2}^\star \xrightarrow{}  \text{SLV}_{i, t_2}(S_{i, t_2}^{\star})$ discussed in \ref{subsection:surrogates}. The assumption states that SLV at $t_2$ is independent of the treatment status $Z_i$, conditionally on the stock value. We use a statistical test to evaluate the plausibility of this assumption. Since the stock value variable is continuous, we first bin it, and test whether the distribution of $\text{SLV}_{i, t_2}(S_{i, t_2}^{\star})$ is the same across the treatment and control groups, conditionally on the binned stock value. We use a two-sample Kolmogorov-Smirnov test, testing that the two samples are drawn from the same distribution under the null. Table \ref{tab:ks} presents the results. None of the tests are rejected at the 95$\%$ confidence level, supporting the surrogacy assumption.

\begin{table}[t]
\small
\centering
\begin{tabular}{llcccc}
\toprule 
\textbf{Stock Value} & $n_\mathcal{C}$ & $n_\mathcal{T}$ & \textbf{KS Statistic} & \textbf{p-value} & \textbf{Reject }$H_0$\\
\midrule
$[0, 10)$ & 2572 & 2671 &0.0189  &0.7282&False \\
\midrule
$[10, 25)$ & 943 & 852 & 0.0189 & 0.996&False\\
\midrule
$[25, 50)$ & 386 & 424 &0.0509 & 0.6461&False\\
\midrule
$[50, 100)$ & 252 & 276 & 0.0751 & 0.4216&False\\
\midrule
$[100, 200)$ & 132 & 144 & 0.0915 & 0.5718&False\\
\midrule
$[200, 500) $ & 109 & 113 & 0.1323 & 0.2545&False\\
\midrule
$[500, 1799]$ & 37 &47  &0.088 & 0.9898&False\\
\bottomrule
\end{tabular}
\caption{Surrogacy checks based on Kolmogorov-Smirnov statistics. }
\label{tab:ks}
\end{table}
\begin{table}[!t]
\centering
\begin{tabular}{lcccc}
\toprule
 & Log Price & Sold Items & Revenue & Profit \\
\midrule
Treatment & $-$0.221$^{***}$ & 1.567$^{***}$ & 53.161$^{***}$ & $-$30.069$^{***}$ \\
 & (0.051) & (0.325) & (14.155) & (6.071) \\[0.5em]
Log Stock Value & 0.055$^{***}$ & 1.243$^{***}$ & 62.801$^{***}$ & $-$14.312$^{**}$ \\
 & (0.013) & (0.295) & (11.677) & (5.756) \\
\midrule
Observations & 8,965 & 8,965 & 8,965 & 8,965 \\
$R^2$ & 0.023 & 0.048 & 0.070 & 0.035 \\
\bottomrule
\multicolumn{5}{l}{\footnotesize $^{*}p<0.1$; $^{**}p<0.05$; $^{***}p<0.01$. Cluster-robust S.E. in parentheses.}
\end{tabular}
\caption{Short-term results of the reactivation experiment: On average, prices decreased by 22\% per item while the number of daily sold items increased by 1.567 units. Revenue  increased on average by around 53 euros per day accompanied by a 30 euro decrease in daily profit per item. Average treatment effect estimates are obtained using OLS panel regression with a difference-in-differences estimator setup. We control for the logarithm of the initial stock value to reduce variance. Displayed absolute treatment effects and standard errors of commercial metrics are scaled with a constant anonymisation factor.}
\label{tab:short_term_results}
\end{table}

Next, we turn to the evaluation of the pricing experiment. Table \ref{tab:short_term_results} provides context on the global short-term commercial ATEs and confirms the intended effect of our pricing intervention. The treatment successfully stimulated demand—sold items and revenue both increased significantly. However, the margin erosion from the 20pp discount increase was severe: despite higher revenue, profit declined substantially, indicating that the aggressive discounting destroyed short term profitability.

The key validation exercise compares the  \textit{forecasted} global long-term impact—captured by SLV efficiency at the end of the experiment—against the \textit{realized} global SLV that materialized 18 months later, once the full product lifecycle had concluded. If the SLV framework correctly accounts for opportunity costs, these two measures should align directionally.

Table~\ref{tab:slv_validation} presents this comparison. Panel A reports raw estimates; Panel B applies 5\% winsorization to address outliers. We include the logarithm of the inital stock value as a control for variance reduction, since initial stock was not balanced across treatment arms and directly affects both SLV efficiency and realized SLV. The treatment effect on SLV efficiency is negative and statistically significant in both panels (column 1), forecasting that the pricing intervention would destroy lifecycle value. The realized SLV measured 18 months after the experiment confirms this prediction: the coefficient is negative in both panels, and statistically significant in the winsorized specification (Panel B, column 2). The realized SLV estimates exhibit substantially higher variance than the SLV efficiency estimates. 
This is due to the forecast uncertainty not accounted for in the construction of the confidence intervals, supporting further the observation made in \cite{duan2021linkedin} that doing so may inflate the Type-I error. This is one short-coming of the current approach, that we plan to address in the next iteration of the project.

Overall, the sign of the effect is consistently negative across all specifications, validating that the SLV framework successfully captured long-term opportunity costs from short-term experimental data in this experiment. Importantly, a business decision to reject this pricing intervention based on the SLV efficiency estimate at the end of the two-week experiment would have been correct — without waiting 18 months to observe the realized outcome.
\begin{table}[!t]
\centering
\vspace{0.5em}
\textit{Panel A: Raw estimates}
\vspace{0.3em}

\begin{tabular}{l r@{\hspace{2cm}} r}
\toprule
 & SLV Efficiency & Realized SLV \\
 & (1) & (2) \\
\midrule
Treatment & $-$36.472$^{***}$ & $-$15.874 \\
 & (11.564) & (72.043) \\[0.5em]
Log Stock Value & $-$17.300 & 982.858$^{***}$ \\
 & (14.854) & (97.292) \\
\midrule
Observations & 8,903 & 8,903 \\
$R^2$ & 0.007 & 0.284 \\
\bottomrule
\end{tabular}

\vspace{1em}
\textit{Panel B: Winsorized estimates (5\%)}
\vspace{0.3em}

\begin{tabular}{l r@{\hspace{2cm}} r}
\toprule
 & SLV Efficiency & Realized SLV \\
 & (1) & (2) \\
\midrule
Treatment & $-$15.845$^{***}$ & $-$44.216$^{***}$ \\
 & (2.291) & (14.094) \\[0.5em]
Log Stock Value & 1.447 & 588.637$^{***}$ \\
 & (1.271) & (8.653) \\
\midrule
Observations & 8,903 & 8,903 \\
$R^2$ & 0.026 & 0.751 \\
\bottomrule
\multicolumn{3}{l}{\footnotesize $^{*}p<0.1$; $^{**}p<0.05$; $^{***}p<0.01$. Cluster-robust S.E. in parentheses.}
\end{tabular}
\caption{Validation of SLV efficiency estimates against realized lifecycle value. Average treatment effect estimates are obtained using OLS regression with difference estimator setup. We control for the logarithm of the initial stock value to reduce variance. Displayed absolute treatment effects and standard errors are scaled with a constant anonymisation factor.}
\label{tab:slv_validation}
\end{table}

\subsection{Customer-based A/B Tests}
\label{subsection:customerslv}

Vouchers and similar promotional instruments are standard tools that accelerate sales, typically achieving an increase in short-term demand and revenue at the cost of reduced profit margins. Standard short-term metrics, which are typically computed over the test window, capture only this immediate effect. But a broad range of customer-level interventions, such as changes to personalization or page ranking algorithms, also significantly impact inventory flow and profitability. For instance, a ranking adjustment might reduce the return rate of top articles but simultaneously slow the sell-through of other stock. This forced future discounting to liquidate the delayed stock creates an illusion of profit in the short-term. Thus, the long-term inventory efficiency of all customer-level campaigns, particularly voucher campaigns, remains an unmeasured factor: whether the accelerated or redirected sales maximized the item's profitability over its remaining lifecycle. We introduce an SLV-based metric adapted to customer-level A/B tests, enabling the measurement of stock efficiency by mirroring the SLV metric definition used in assortment-based A/B tests.

For a test running between two dates $t_1$ and $t_2$, we define the customer-level SLV efficiency for a customer $c$ and unit $i$ as
$$
\widehat{\text{SLVeff}}_{i,c} = \sum_{s=t_1}^{t_2} (\text{MP}_{i, c, s} - \widehat{\text{nSLV}}_{i, s} \cdot N_{i,c,s} \cdot \text{pp}_i)\,,
$$
where $\text{MP}_{i, c, s}$ is the marginal profit generated by customer $c$ on date $s$ for unit $i$, 
and $N_{i,c,s}$ denotes the number of units $i$ bought by $c$ on day $s$. All terms are equal to 0 if no purchase was made. $\widehat{\text{SLVeff}}_{i,c}$ represents the difference between the instant marginal profit realised on a given date, and the average marginal profit that could have been made selling this article after that date, until the end of its lifecycle. It generalises the SLV efficiency metric defined in section \ref{subsection:slvefficiency} at the unit level, by linking the purchases to the customer who bought it, see Appendix \ref{appendix:slvcustomers} for more details. In a customer-level A/B test, we report the average difference of  $\widehat{\text{SLVeff}}_{i,c}$ over the treatment and control groups of customers, across the whole assortment. The sign of the difference informs about the stock efficiency of the customer-level intervention.

\paragraph{Example 1}
\label{section:example1vouchers}
In a voucher experiment, suppose that all customers in the treatment group receive a voucher, and the control group receives none. We anticipate a sharp increase in short-term revenue, accompanied by a decrease in instantaneous profit margins. Reporting the results on SLV efficiency brings a more nuanced answer by incorporating the opportunity cost of the stock. For items $i$ with low nSLV (e.g., slow sellers requiring high future discounting), the difference $\text{MP}_{i, c, s} - \widehat{\text{nSLV}}_{i, s} \cdot N_{i,c,s} \cdot \text{pp}_i$ may be less negative than expected. This indicates that the profit loss, when viewed against the item's remaining profitability, is not as severe as suggested by the short-term profit outcome alone. Furthermore, the nSLV forecasts can be utilized preemptively for voucher creation, allowing us to identify and target eligible units—specifically those at high risk of terminal discounting—for immediate redemption.

\paragraph{Example 2}
\label{section:example2ranking}
Consider a ranking algorithm experiment where customers in the treatment group are promoted with ``fast-moving" articles (bestsellers) to maximize immediate conversion. Although this strategy yields high short-term marginal profit, it often cannibalizes visibility for ``at-risk" inventory. Because bestsellers have a high nSLV, the treated customers are likely to have a smaller SLV efficiency than the control customers, leading to a negative effect between the two groups. This negative value is the critical insight provided by the SLV metric: it exposes the true inefficiency where the policy promotes high-value units while creating overstock that requires deeper discounts at the  end of the season.

\subsection{Optimal algorithmic discounting}
\label{subsection:optimal_discounting}
Zalando's pricing system follows a forecast-then-optimize architecture \citep{birr2025}. A demand forecasting model generates predictions for different discount levels and market conditions, producing a discount-time grid that captures expected sales across pricing scenarios. This forecasting output feeds into an optimizer that selects the discount level for each article that maximizes a specified objective function.
The pricing objective must navigate a fundamental tension. Revenue optimization favors higher discounts and faster stock clearance, while profit optimization may indicate that selling at a later time with a lower discount is preferable. To address this, the system employs a multi-objective optimization framework using the weighted sum method \citep{Ehrgott2005}.

The optimization target for the pricing algorithm is formulated as follows:
For some steering parameter $\alpha \in [0, \infty)$, for articles $i \in N$, and discounts $d_i
  \in D$, our optimization objective is
\begin{equation}\label{eq:optProblem}
  \max_{d_{i}\in D} \ \text{LTP}_i(d_i) + \alpha \cdot \text{NMV}_i(d_i)  \ \forall
  i
  \in N.
\end{equation}
As established in optimization theory, any solution maximizing this objective will be Pareto-efficient \citep{marler2010weighted}. The parameter $\alpha$
controls the trade-off between long term profit (LTP) and Net Merchandise Value (NMV), with higher values prioritizing short-term revenue. Stakeholders can evaluate offers across multiple $\alpha$ values to select pricing strategies aligned with their business objectives. We employ the standard approach of identifying Pareto-efficient offers \citep{Ehrgott2005}—offers that cannot be dominated by competing alternatives in both NMV and LTP dimensions. However, since many Pareto-efficient offers typically exist, expert stakeholder knowledge is essential for final selection.

To compute this objective, the system requires estimates of both NMV and LTP for each article-discount combination. NMV is calculated from predicted sales $s_{d}$, the discount level $d$, the undiscounted price $P$, the predicted return rate $R_{d}$, and the applicable value-added tax VAT:

\begin{equation}\label{eq:nmv}
  \text{NMV}(d) = (1-R_{d}) \cdot \frac{(1-d) \cdot s_d \cdot P}{(1 + \text{VAT})}.
\end{equation}

For long-term profit, the system computes both short-term profit from sales during the pricing period and the future profit potential of unsold inventory.
Short-term profit is given by:
\begin{equation}\label{eq:p}
  p(d) = \text{NMV}(d) - C \cdot s_d.
\end{equation} where $C$ represents fulfillment and return costs. The future profit potential of remaining stock is captured through an opportunity cost term $\phi(d)$, leading to the full LTP calculation:

\begin{equation}\label{eq:ltp}
  \text{LTP}(d) = p(d) + \underbrace{(M - s_{d} (1-R_{d})) \cdot \text{RV}}_{\phi(d)}.
\end{equation}
Here $M$ denotes current stock and $RV$ represents the per unit expected future profit for the article, its residual value. Each article has its own $RV$ that evolves over the pricing period. 

The optimization framework assumes independence across articles: the discount applied to one article does not affect the sales projections of another. This separability extends to the objective function in Equation \eqref{eq:optProblem}, which contains no cross-article terms. As a result, NMV and LTP can be computed independently for each article-discount combination, and optimal discounts determined by evaluating these values per article in isolation. This structure permits straightforward parallelization, allowing the full assortment pricing problem to be solved efficiently at scale. Please see \cite{birr2025} for a more thorough discussion of the current algorithmic pricing system, the forecasting framework and the $RV$ calculation in production. \cite{Huelden2024} and \cite{kunz2023deep} provide further insights into the pricing systems at Zalando.

The conceptual challenges in this optimization problem are threefold. The optimizer needs the best performing demand and return rate forecasts $s_{d}$ and $R_{d}$ and the most accurate $RV$ representation. Here, the SLV framework offers a natural solution. $RV$ can be obtained by the forecasted $nSLV$ per unit of remaining stock, conditional on current stock levels, time remaining in season, and other relevant state variables, derived from the forecast introduced in section \ref{section:rvforecast}.  

This formulation preserves the structure of the existing optimization framework while providing several advantages. First, nSLV forecasts are derived from the same modeling framework used elsewhere in the SLV methodology, ensuring consistency in assumptions and feature definitions across the company and measurement domain. 
Second, nSLV forecasts can be conditioned on a flexible set of state variables tailored to the pricing context. Beyond article identity and time, nSLV can incorporate current stock levels, recent sell-through velocity, and other signals that affect future profit potential. This flexibility allows the future value estimate to reflect the specific information available at decision time.
Third, using nSLV creates consistency between measurement and optimization. When evaluating pricing A/B tests using SLV (as described in Section \ref{subsection:slvefficiency}), the same forecasts that assess the treatment effects can serve as input to the optimizer. This alignment ensures that the metric used to judge success is the same quantity the algorithm optimizes, avoiding the potential for divergence between the evaluation criteria and the decision rules.

\subsection{Annualisation}
\label{subsection:annualization}
Online experiments typically run for days or weeks, yet the effects of interest—on customer lifetime value, brand equity, or sustainable profitability—accumulate over months or years.  Organizations require credible annualized estimates of treatment effects to support investment decisions, resource allocation, and performance reporting. Once A/B tests have been analysed and new improvements launched, organizations need to translate short term experimental results into annualized business impacts to enable dollar-based cost-benefit analysis and consistent tracking across experimentation programs \citep{netflixtechblog2025hte}. In this section, we develop a framework for translating short term experimental results into annualized business impact estimates. Our approach leverages the SLV efficiency relative metric introduced in Section \ref{subsection:slvefficiency}, which captures long term profit implications within short experimental windows, as the basis for principled extrapolation to annual effects.

The pricing A/B tests we seek to annualize at Zalando typically run for one to four weeks and evaluate new pricing policies against a business-as-usual control. The tested intervention usually involve changes to the algorithmic pricing system described in Section \ref{subsection:optimal_discounting}. These include improvements to input forecasts (such as return rate or demand predictions), modifications to the optimization objective or constraints, and rollouts of entirely new pricing algorithms as in \cite{kunz2023deep} or \cite{birr2025}. Other A/B tests investigate the effect of new pricing policies such as human-price interventions as discussed in \cite{Huelden2024}. Each of these interventions affects markdown decisions across the assortment and generates treatment effects that must be translated into annual business impact.

Suppose we run such a short-term experiment over two weeks, testing a new pricing algorithm and obtain a treatment effect on relative and absolute short term metrics such as revenue or profit as well as on $\widehat{\text{relSLVeff}}$. 
For seasonal assortment in fashion e-commerce, the central challenge in annualizing experimental results is that absolute metrics measured during the experiment depend on stock levels at the period start. These initial stock levels vary throughout the year. Seasonal stock operates under fixed inventory constraints: articles are purchased at season start, and the only path to increased revenue and profit is selling this fixed stock more efficiently over the product lifecycle. A short term experiment captures efficiency gains on the available inventory during the test window, but the absolute effect does not directly extrapolate to an annual impact. 
The relative SLV efficiency metric provides a solution to this problem. As a normalized measure of stock efficiency, it captures the percentage improvement in value extracted from consumed inventory, independent of the absolute stock levels. Next, we explain how to annualise estimates on both Gross Merchandising Revenue (GMV) and profit, based on $\widehat{\text{relSLVeff}}$.

\paragraph{Annualizing Incremental Revenue.} 
Throughout our framework, we rely on our internal business accounting logic to define the commercial performance metrics and on simplifying assumptions in order to produce actionable insights in a complex commercial environment. Our first and core assumption in our framework is relevant for both revenue and profit annualization:

\newtheorem{assumption}{Assumption}

\begin{assumption}\label{ass:constant_slv}
The relative SLV efficiency based on marginal profit remains constant over the year.
\end{assumption}

\noindent Under assumption \ref{ass:constant_slv}, the annualized incremental revenue is:
\begin{equation}\label{eq:annualized_gmv}
AnnualizedIncr_{GMV} = \widehat{\text{relSLVeff}} \times \text{Baseline GMV (eligible assortment)} \times Risk Factor,
\end{equation}
where  $\text{Baseline GMV (eligible assortment)}$ is the total GMV of seasonal stock, eligible for algorithmic pricing,  sold in the 12 months preceding the annualization date. It is taken as a reference baseline adapted to the eligible assortment for annualization and can be adjusted for (market specific) projected growth rates to reflect growth expectations. The $Risk Factor$ addresses potential overestimation from dilution, winner's bias, and seasonal aspects not fully captured by SLV efficiency. We adopt a conservative risk factor lower than 1 to all annualized estimates. This accounts for the combined uncertainty from the factors discussed above and aligns with annualization practices used elsewhere in the organization. 

\begin{figure}[t]
\centering
\includegraphics[width=1\linewidth]{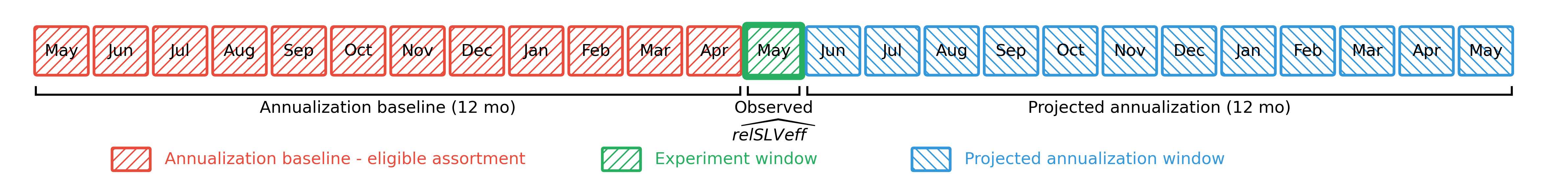}
\caption{Visualized annualization framework for a exemplary short term experiment conducted in May. The baseline revenue is computed from the 12 months preceding the experiment from the stock eligible to be discounted.}
\label{fig:slv_annualization_diagram}
\end{figure}

Figure \ref{fig:slv_annualization_diagram} illustrates the annualization approach. Note that we use the relative SLV efficiency estimate based on marginal profit to annualise GMV. This is justified from a simple approximation, where the details are provided in Appendix \ref{appendix:annualisation}. An approximation is used for simplicity, since stakeholders are generally interested in getting a correct order of magnitude, rather than a precise estimate.

\paragraph{Annualizing Incremental Profit.} The annualization of profit requires assumptions about how costs behave under different pricing policies. 

\begin{assumption}\label{ass:fixed_stock}
The stock is given, i.e., purchasing prices have already materialized.
\end{assumption}

\begin{assumption}\label{ass:cost_independence}
Logistics, payment and storage costs are independent of the pricing policy.
\end{assumption}

\noindent Assumption \ref{ass:fixed_stock} reflects the nature of seasonal inventory: goods are purchased at the start of the season, and the pricing policy operates on this fixed stock. The assumption holds provided that within-season reorders are negligible relative to initial inventory—a reasonable approximation for most seasonal fashion assortments at Zalando. This may slightly underestimate long-term effects if the improved pricing performance leads to larger purchasing volumes in subsequent seasons.

\begin{figure}[t]
\centering
\includegraphics[width=.75\linewidth]{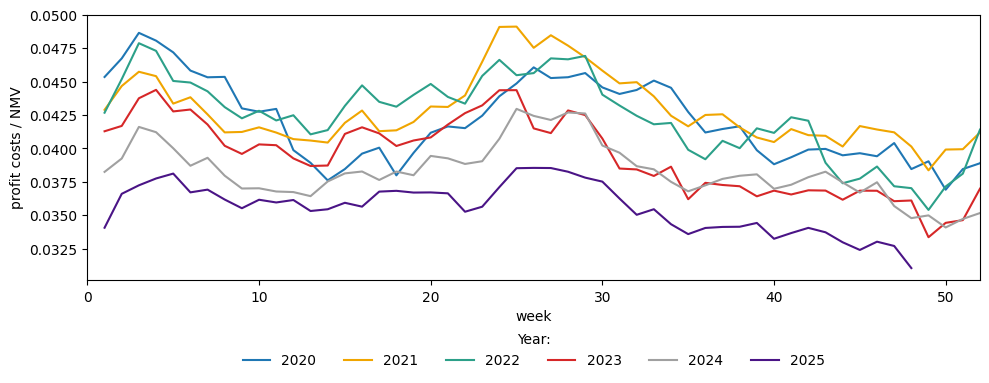}
\caption{Average profit cost share of revenue across calendar weeks in the past 5 years. Displayed values are multiplied with a constant anonymisation factor.}
\label{fig:pcii_cost_share_week_year}
\end{figure}

Assumption \ref{ass:cost_independence} states that pricing policy changes do not meaningfully affect fulfillment, logistics, or other variable costs captured in the overall profit. Empirical evidence supports this assumption: Figure \ref{fig:pcii_cost_share_week_year} displays the ratio of profit costs to NMV. The cost share remains approximately constant year-over-year, varying by only 2--4 percentage points across different pricing regimes including black price periods and end-of-season sales during the year.

Under Assumptions \ref{ass:fixed_stock} and \ref{ass:cost_independence}, the relationship between profit and revenue simplifies considerably. By definition, $\text{Profit} = \text{MP} - \text{pp}$ and $\text{MP} = \text{NMV} - \text{IncrementalCosts}$. Under Assumption \ref{ass:fixed_stock}, the incremental profit from a new pricing policy is $\Delta \text{Profit} = \Delta \text{MP}$ since $\Delta \text{pp} = 0$. Under Assumption \ref{ass:cost_independence}, $\Delta \text{MP} = \Delta \text{NMV}$ since $\Delta \text{IncrementalCosts} = 0$. Combining these results with Assumption \ref{ass:constant_slv}:
\begin{equation}\label{eq:annualized_pcii}
AnnualizedIncr_{profit} = AnnualizedIncr_{NMV} = (1-\text{VAT}) \cdot AnnualizedIncr_{GMV}.
\end{equation}

\noindent This means that for the seasonal stock, the annualized profit can be derived directly from the annualized GMV by converting to NMV (adjusting for VAT), without requiring a separate profit-based calculations. 

\begin{figure}[t]
\centering
\includegraphics[width=.75\linewidth]{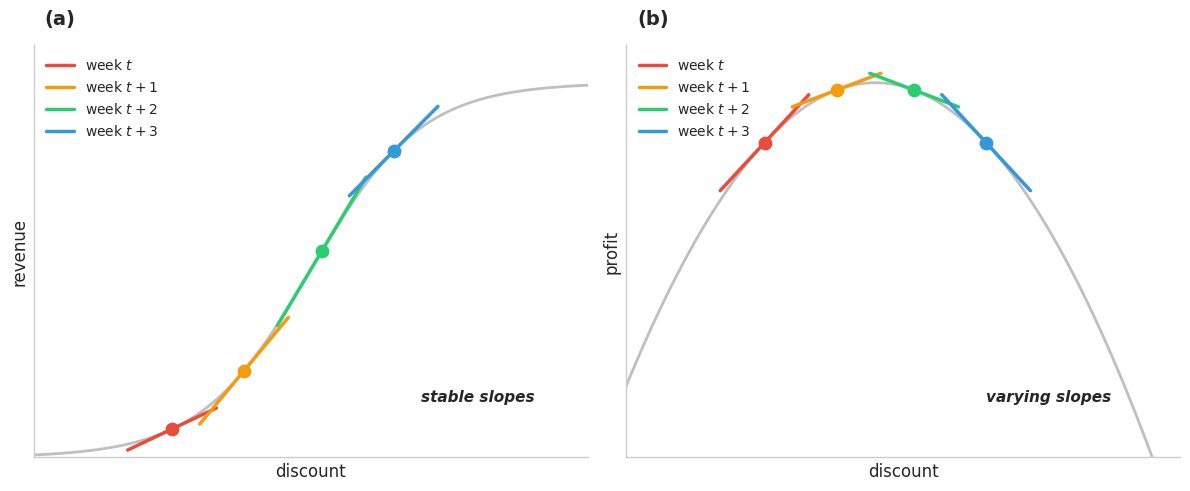}
\caption{Stability of the relationship between pricing strategy and revenue or profit. Colored lines show the marginal effect at each week's discount level. (a) Revenue follows an S-shaped response; despite this nonlinearity, the marginal effect remains relatively stable across operating points. (b) Profit is concave in discount; the marginal effect varies from positive to negative as the discount regime shifts, making profit-based measures sensitive to measurement timing.}
\label{fig:revenue_profit_stability}
\end{figure}

A natural question is why using relative SLV efficiency based on marginal profit instead of directly using a profit-based version to annualize profit effects. The answer lies in the stability of the underlying functional relationships. Revenue is a monotonic, approximately linear function of discount: deeper discounts yield proportionally higher sales volume and revenue. Profit, by contrast, is a concave function of discount—it initially increases as discounts stimulate demand, reaches a maximum, and then decreases as margin erosion outpaces volume gains. Figure~\ref{fig:revenue_profit_stability} illustrates this distinction. The revenue-discount curves shift vertically or horizontally across weeks as demand conditions change, but their slopes remain relatively stable. The profit-discount curves, however, not only shift but also change shape: the slope at any given discount level varies substantially depending on when in the season the measurement occurs. Assumption \ref{ass:constant_slv} for marginal profit—that the relative efficiency gain on revenue remains constant—is therefore robust to these week-over-week shifts. Assuming that the relative efficiency gain on profit remains constant, however, requires the more stringent condition that we measure at the same point on the concave profit curve each week, which is unlikely to hold in practice.

The annualisation framework presented in this section is based on the SLV efficiency relative estimate. Although SLV efficiency relative can be defined for businesses that operate under stock constraints, our recommendation is that practitioners must identify and validate a set of underlying operating assumptions, analogous to Assumptions \ref{ass:fixed_stock} and \ref{ass:cost_independence}, that accurately reflect the specific inventory dynamics and cost structure of their unique business model.

We close this chapter with an illustration of the annualization process for selected real world A/B tests run at Zalando.
The experiments tested the impact of an improved return rate forecast on the commercial performance of our main pricing algorithm described here \cite{birr2025}.
The forecast is relevant in the algorithmic formulation of (\ref{eq:nmv}) and (\ref{eq:ltp}) which influences the quality of the optimization function (\ref{eq:optProblem}) as outlined in section \ref{subsection:optimal_discounting}. The central innovation of the return forecast improvement was the improvement of the return elasticity with respect to the discount. That enabled a more precise forecast of the return rate.

\begin{table}[!h]
\small
\centering
\begin{tabular}{llccc}
\toprule 
\textbf{Experiment} & \textbf{Metric} & \textbf{SLV efficiency relative} & \textbf{Baseline Value} & \textbf{Annualised Value} \\
\midrule
1 & Marginal Profit & 0.9\% & & \\
  & Revenue         &       & 1331 Mio Eur & 8.38 Mio Eur \\
  & Profit          &       & 192 Mio Eur  & 6.71 Mio Eur \\
\midrule
2 & Marginal Profit & 0.7\% & & \\
  & Revenue         &       & 1453 Mio Eur & 7.11 Mio Eur \\
  & Profit          &       & 207 Mio Eur  & 5.69 Mio Eur \\
\midrule
\midrule
\textbf{Average} & Marginal Profit & \textbf{0.8\%} & & \\
                 & Revenue         &       &  & \textbf{7.75 Mio Eur} \\
                 & Profit          &       &  & \textbf{6.20 Mio Eur} \\
\bottomrule
\end{tabular}
\caption{Annualization Results for the Return Forecast Pricing A/B tests. Displayed euro values are multiplied with a constant anonymisation factor, keeping true values confidential.}
\label{tab:annualization}
\end{table}

During a first 3-week experiment in 24 markets in Europe affecting over 4 million articles, we measured the commercial impact of the algorithmic intervention. The outcomes were positive and statistically significant with a measured 0.9\% global uplift in relative SLV efficiency. Table \ref{tab:annualization} summarizes the results of the annualization calculation. All numbers are scaled by an anonymisation factor to keep the internally reported values confidential.
Based on this first test, we project an annual 8.38 Mio Eur uplift in revenue accompanied by a 6.71 Mio Eur uplift in profit.
As stated in Assumption \ref{ass:constant_slv}, the core assumption in the annualization of experimental effects is a constant effect of relative SLV efficiency throughout the year. In order to test the validity of this requirement, we ran a second identical return forecast experiment 3 months after the first one concluded. The test ran globally in 24 markets for 3 weeks on around 6 million articles. We measured a similar uplift in relative SLV efficiency of 0.7\% globally and calculated a projected annual 7.11 Mio Eur uplift in revenue accompanied by a 5.69 Mio Eur uplift in profit.
Averaging over both A/B tests gives us thus an overall annual uplift of 7.75 Mio Eur revenue and 6.20 Mio Eur of incremental profit after rollout of the new intervention. After review with the finance and pricing department, the algorithmic intervention was subsequently approved, prioritized for production and subsequently rolled-out.
\section{Conclusion}
\label{section:conclusion}

This paper introduced Stock Lifetime Value as a framework for measuring the long term opportunity cost of commercial interventions in e-commerce experimentation. The core contribution is the SLV efficiency metric, which enables reliable long term inference from short experimental windows by comparing realized short term profit against the forecasted opportunity cost of consumed inventory. We empirically validated this approach with an A/B test, showing that SLV efficiency estimates from a two-week experiment correctly predicted the loss in profit that would have been observed 18 months later.

Beyond measurement, we discussed three practical applications of the framework. First, we extended SLV efficiency to customer-based A/B tests, enabling evaluation of voucher campaigns and ranking algorithms against their true inventory efficiency. Second, we showed how SLV integrates naturally as an optimization target for pricing algorithms, replacing ad-hoc future value estimates with a unified metric that aligns measurement and optimization. Third, we developed an annualization methodology that translates short term experimental results into financial reporting metrics required by business stakeholders.

The framework rests on several assumptions that practitioners should validate for their specific context: structural stationarity in data and pricing processes, the surrogacy assumption that short term profit and stock levels at the end of the experiment fully mediate long term outcomes, and stability in cost structures for the annualization approach. Future work could extend the approach to settings with significant within-season restocking. Another crucial aspect moving forward is to take the forecast uncertainty into account when constructing confidence intervals for SLV efficiency. This would reduce the type-I error on this metric as indicated by \cite{duan2021linkedin}.

\bibliographystyle{alpha}
\bibliography{arxive_sample}

\appendix

\clearpage

\clearpage

\section{SLV calculation}
\label{appendix:slv}

This section illustrates on a toy example how $\text{SLV}_{i, t_0}(S_{i,t_0})$ is computed for an item $i$, given that we have $S_{i,t_0}=10$ units in stock at time $t_0$. Consider the following data on stock levels, returns from customers, items sold, and the assolicated daily marpinal profit.

\begin{table}[!h]
\small
\centering
\begin{tabular}{l|c|c|c|c|c}
\toprule 
Date & $t_0$ & $t_0+1$ & $t_0+2$ & $t_0+3$ & $t_0+4$ \\
\midrule
Stock & 10 & 7 & 5 & 3 & 0\\
\midrule
Returns  & 2 & 0 & 1 & 0 & 0\\
\midrule
Items Sold & 5 & 2 & 3 & 3 & 0\\
\midrule
Marginal Profit  &  40  & 20   & 25 & 25 &  0\\
\bottomrule
\end{tabular}
\end{table}

\noindent Returns that materialise on a date $t$ are from purchases made prior to $t$. The marginal profit of items returned on a given date are retrospectively updated one the return has materialised, and associated with a negative marginal profit for that sale, due to logistics and and payment costs. With this data, we have $\text{SLV}_{i, t_0}(S_{i,t_0}) = 40 + 20 + 25 = 85$. At time $t_0+1$, we have 7 units in stock, $S_{i,t_0+1}=7$. The associated SLV is the cumulative profit of the 2 items sold at $t_0+1$, the 3 items sold at time $t_0+2$, and of the 2 out of 3 items sold at $t_0+3$, giving $\text{SLV}_{i, t_0+1}(S_{i,t_0+1}) = 20 + 25 + 2*25/3 = 61.67$. 

\section{SLV efficiency for customers}
\label{appendix:slvcustomers}

For an article $i$ in group $j\in\{0,1\}$ in an A/B test between $t_1$ and $t_2$, SLV efficiency at the article level is defined in (\ref{eq:slveff}) as
$$
\text{SLVeff}_i(S_{i, t_1})(j) = \sum_{t=t_1}^{t_2} \text{MP}_{i,t} \mathbf{1}\left(\sum_{s=t_0}^t N_{i,s} \leq S_{i,t_1}\right) - \text{SLVcons}_{i, t_1, t_2}(j)\,,
$$
where $S_{i, t_1}$ denotes the number of articles $i$ in stock at $t_1$, and
$$
\text{SLVcons}_{i, t_1, t_2}(j) = \text{SLV}_{i, t_1}(S_{i, t_1})(0,0)-\text{SLV}_{i, t_2}(S_{i, t_2}^{\star})(j,0)
$$
quantifies the opportunity costs of the consumed inventory in during the A/B test period. We identify two assumptions under which $\text{SLVcons}_{i, t_1, t_2}$ simplifies, allowing a simple adaptation of the metric at customer level.\\

\noindent\textbf{Assumption A: Constant nSLV.} $\text{nSLV}_{i, t_1}(0,0) = \text{nSLV}_{i, t_2}(0,0)= \text{nSLV}_{i, t_2}(1,0)$.\\

\noindent\textbf{Assumption B: Incorporating Returns and Restocking.} We assume that the simplified calculation incorporates sales from non-anticipated returns and non-anticipated restocking events during the experiment period.\\

Under Assumption B, the difference in stock between $t_1$ and $t_2$ is the number of sales that occured in the A/B test,
$$
S_{i,t_1}-S_{i,t_2}^\star = S_{i,t_1} - max(0, S_{i, t_1} - \sum_{t=t_1}^{t_2} N_{i,t}) = \sum_{t=t_1}^{t_2} N_{i,t}\,.
$$
Together with Assumption A, this yields
\begin{align}
\text{SLVcons}_{i, t_1, t_2}(j) &= \text{nSLV}_{i, t_1}(0,0)\cdot \text{pp}_i\cdot S_{i,t_1}-\text{nSLV}_{i, t_2}(j,0)\cdot \text{pp}_i\cdot S_{i,t_2}^\star\notag\\
&= \text{nSLV}_{i, t_1}(0,0)\cdot \text{pp}_i \cdot \sum_{t=t_1}^{t_2} N_{i,t}\,.\notag
\end{align}
and
$$
\text{SLVeff}_i(S_{i, t_1}) = \sum_{t=t_1}^{t_2} (\text{MP}_{i,t}  - \text{nSLV}_{i, t_1}(0,0)\cdot \text{pp}_i \cdot N_{i,t})\,.
$$
Let $\text{MP}_{i, c, t}$ be the marginal profit generated by customer $c$ on date $t$ for unit $i$, 
and $N_{i,c,t}$ the number of units $i$ bought by $c$ on day $t$. Then
$$
\text{SLVeff}_i(S_{i, t_1}) = \sum_c \sum_{t=t_1}^{t_2} (\text{MP}_{i,c,t}  - \text{nSLV}_{i, t_1}(0,0)\cdot \text{pp}_i \cdot N_{i,c,t}) =: \sum_c \text{SLVeff}_{i, c}\,.
$$
The customer-level $\text{SLVeff}_{i, c}$ defined in Section \ref{subsection:customerslv} relaxes the assumption of a constant nSLV, and uses $\text{nSLV}_{i, t}(0,0)$ instead of $\text{nSLV}_{i, t_1}(0,0)$ in the last written expression. 

\section{Annualisation}
\label{appendix:annualisation}
\begin{figure}[t]
\centering
\includegraphics[width=.75\linewidth]{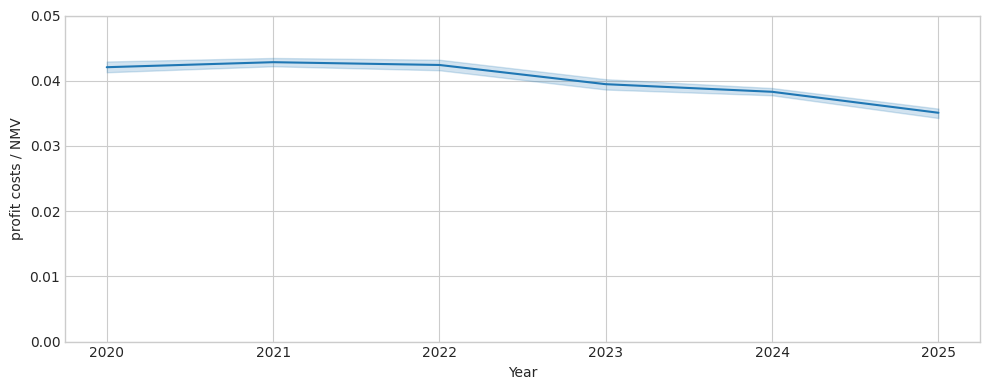}
\caption{Average profit cost share of revenue in the past 5 years. Displayed values are multiplied with a constant anonymisation factor.}
\label{fig:pcii_cost_share_yearly}
\centering
\includegraphics[width=.75\linewidth]{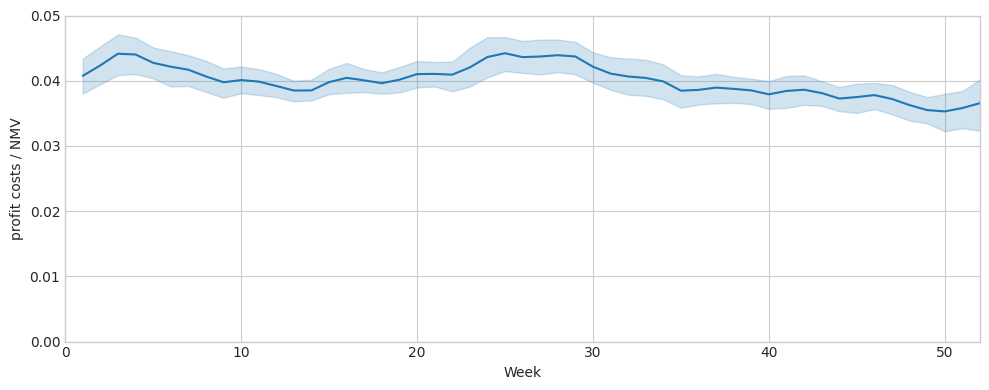}
\caption{Average profit cost share of revenue across calendar weeks in the past 5 years. Displayed values are multiplied with a constant anonymisation factor.}
\label{fig:pcii_cost_share}
\end{figure}

In section \ref{subsection:annualization}, we use the relative SLV efficiency based on marginal profit to annualize GMV, rather than defining a GMV-based SLV efficiency metric directly. Motivated by the definition of $\widehat{\text{relSLVeff}}^{MP}$, where the superscript highlights that MP is used as a building block for the definition of SLV, we can construct:
\begin{equation*}
\widehat{\text{relSLVeff}}^{GMV} = \frac{\overline{\text{GMV}}_{trt} - \overline{\text{GMV}}_{ctl} - \Delta\widehat{\text{SLVcons}}^{GMV}}{\overline{\text{GMV}}_{ctl} + \Delta\widehat{\text{SLVcons}}^{GMV}},
\end{equation*}
where $\overline{\text{GMV}}_{trt}$ and $\overline{\text{GMV}}_{ctl}$ denote the average stock-constrained short-term GMV sold in the treatment and control groups during the experiment, and $\Delta\widehat{\text{SLVcons}}^{GMV}$ reflects the difference in SLV consumed between treatment and control on the GMV scale. Forecasting $\Delta\widehat{\text{SLVcons}}^{GMV}$ would require training a separate nSLV model based on cumulative GMV rather than marginal profit from historical sales. However, a simple approximation obviates this need. The accounting relationship between marginal profit and GMV is given by:
\begin{equation*}
\text{MP} = \text{NMV} - \text{Profit costs} = (1 - c) \cdot \text{NMV} = (1 - c) \cdot \frac{1}{1 + \text{VAT}} \cdot \text{GMV} = \kappa \cdot \text{GMV},
\end{equation*}
where $c$ denotes the incremental cost share as a fraction of NMV, VAT is the average value-added tax rate across European markets, and $\kappa = (1-c)/(1 + \text{VAT})$ is the resulting proportionality constant. Figure \ref{fig:pcii_cost_share_yearly} displays the average profit cost share over the past 5 years. Figure \ref{fig:pcii_cost_share} displays the within-year variation between 2020 and 2025. While the incremental cost share varies considerably across individual articles, for any given article the ratio tends to converge to a stable value over time as sales accumulate. Moreover, because the assortment composition remains similar year-over-year—with comparable mixes of categories, price points, and sourcing structures—the aggregate cost share across the full assortment stays approximately constant. This proportionality implies:
\begin{equation*}
\Delta\widehat{\text{SLVcons}}^{GMV} \approx \Delta\widehat{\text{SLVcons}}^{MP} / \kappa.
\end{equation*}
Since the same scaling factor $\kappa$ applies to both the short-term realized values ($\overline{\text{GMV}}$ and $\overline{\text{MP}}$) and the SLV consumed terms, it cancels in the ratio, yielding $\widehat{\text{relSLVeff}}^{GMV} \approx \widehat{\text{relSLVeff}}^{MP}$. This allows us to use the existing marginal profit-based metric directly for GMV annualization without introducing additional modeling complexity.

\end{document}